\documentclass[preprint]{aastex}

\shorttitle{Giant HII Regions. V. G333.1--0.4}
\shortauthors{Figuer\^ edo et al.}

\usepackage[dvips]{color}

\definecolor{darkred}{rgb}{0.6,0.0,0.0}

\begin{document}

\title{The Stellar Content of Obscured Galactic Giant HII Regions V:
G333.1--0.4}

\author{E. Figuer\^ edo\altaffilmark{1}}
\affil{IAG--USP, R. do Mat\~ao 1226, 05508--900, S\~ ao Paulo, Brazil} 
\email{lys@astro.iag.usp.br}

\author{R. D. Blum\altaffilmark{1}}
\affil{Cerro Tololo Interamerican Observatory, Casilla 603, La Serena, Chile} 
\email{rblum@noao.edu}

\author{A. Damineli\altaffilmark{1}}
\affil{IAG--USP, R. do Mat\~ao 1226, 05508--900, S\~ ao Paulo, Brazil}
\email{damineli@astro.iag.usp.br}

\and

\author{P. S. Conti}
\affil{JILA, University of Colorado \\ Campus Box 440, Boulder, CO, 80309}
\email{pconti@jila.colorado.edu}

\altaffiltext{1}{Visiting Astronomer, Cerro Tololo Inter--American Observatory,
National Optical Astronomy Observatories, which is operated by Associated
Universities for Research in Astronomy, Inc., under cooperative agreement with
the National Science Foundation.}

\begin{abstract}
We present high angular resolution near--infrared images of the
obscured Galactic Giant HII (GHII) region G333.1--0.4 in which we
detect an OB star cluster. For G333.1--0.4, we find OB stars and other
massive objects in very early evolutionary stages, possibly still
accreting. We obtained $K$--band spectra of three stars; two show O
type photospheric features, while the third has no photospheric
features but does show CO 2.3 $\mu$m band--head emission. This object
is at least as hot as an early B type star based on its intrinsic
luminosity and is surrounded by a circumstellar disc/envelope which
produces near infrared excess emission. A number of other relatively
bright cluster members also display excess emission in the $K$--band,
indicative of disks/envelopes around young massive stars. Based upon
the O star photometry and spectroscopy, the distance to the cluster is
2.6 $\pm$ 0.4 kpc, similar to a recently derived kinematic (near side)
value. The slope of the $K$--band luminosity function is similar to
those found in other young clusters. The mass function slope is more
uncertain, and we find $-1.3 \pm 0.2 < \Gamma < -1.1 \pm 0.2$- for
stars with M $> 5$ M$_\odot$ where the upper an lower limits are
calculated independently for different assumptions regarding the
excess emission of the individual massive stars. The number of Lyman
continuum photons derived from the contribution of all massive stars
in the cluster is $0.2$ $\times$ $10^{50}$ $s^{-1}$ $< NLyc < 1.9$
$\times$ $10^{50}$ $s^{-1}$.  The integrated cluster mass is $1.0$
$\times$ $10^{3}$ $M_\odot < M_{cluster} < 1.3$ $\times$ $10^{3}$
$M_\odot$.

\end{abstract}

\keywords{HII regions --- infrared: stars --- stars: early-type ---
stars: fundamental parameters --- stars: formation}

\section{Introduction}

Massive stars have a strong impact on the evolution of
Galaxies. O--type stars and their descendants, the Wolf--Rayet stars,
are the main source of UV photons, mass, energy and momentum to the
interstellar medium. They play the main role in the ionization of the
interstellar medium and dust heating. The Milky Way is the nearest
place to study, simultaneously, massive stellar populations and their
impact on the surrounding gas and dust. The sun's position in the
Galactic plane, however, produces a heavy obscuration in the visual
window ($A_V \approx 20 - 40$ mag) toward the inner Galaxy, where
massive star formation activity is the largest. At longer wavelengths
such as in the near infrared, the effect of interstellar extinction is
lessened ($A_K \approx 2 - 4$ mag), yet the wavelengths are still
short enough to probe the stellar photospheric features of massive
stars \citep{han96}.

The study of Giant HII regions (GHII -- emitting at least
$10^{50}$~LyC photons s$^{-1}$, or $\approx$~10 $\times$ Orion) in the
near--infrared can address important astrophysical issues such as:
1. characterizing the stellar content by deriving the initial mass
function (IMF), star formation rate and age; 2. determining the
physical processes involved in the formation of massive stars, through
the identification of OB stars in very early evolutionary stages, such
as embedded young stellar objects (YSOs) and ultra--compact HII
regions (UCHII); and 3. tracing the spiral arms of the Galaxy by
measuring spectroscopic parallaxes for main sequence OB stars. The
exploration of the stellar content of obscured Galactic GHII regions
has been studied recently by several groups: \citet{han97},
\citet[2000, 2001]{blum99}, \citet{fig02} and \citet{okum00}.  These
observations revealed massive star clusters at the center of the HII
regions which had been previously discovered and studied only at much
longer radio wavelengths.

In this work, we present results for G333.1--0.4 (R.A.$=$ 16h21m03.3s
and DEC. $=$ $-$50d36m19s J2000), located at a kinematic distance 2.8
kpc (near side) or 11.3 kpc (far side), which we adopted from
\citet{vil00} with R$_0$ = 7.9 kpc. Inward of the solar circle the
galactic kinematic rotation models give two values for the distance. A
difficulty with such models comes from the classical two--fold
distance distance ambiguity for lines of sight close to the Galactic
Center (GC) \citep{wat03}. Furthermore, non--circular velocity
components can lead to erroneous distances. We will show below that a
spectroscopic parallax method leads to a distance of 2.6 $\pm$ 0.4
kpc. G333.1--0.4 does not appear in visible passband images, but in
the infrared one sees a spectacular star formation region.

In the present paper, we present an investigation of the stellar
content of G333.1--0.4 through the $J$, $H$ and $K$ imaging and
$K$--band spectroscopy (described in \S2). In \S3 we consider the
photometry, and in \S4 we analyze the spectra. We determine the
distance in \S5 and discuss the results in \S6.

\section{Observations and Data Reduction}

$J$ ($\lambda$ $\approx$ 1.3 $\mu$m, $\Delta$$\lambda$ $\approx$ 0.3
$\mu$m), $H$ ($\lambda$ $\approx$ 1.6 $\mu$m, $\Delta\lambda$
$\approx$ 0.3 $\mu$m) and $K$ ($\lambda$ $\approx$ 2.1 $\mu$m,
$\Delta$$\lambda$ $\approx$ 0.4 $\mu$m) images of G333.1--0.4 were
obtained on the night of 1999 May 1 and a new set of images 65$''$
east of the cluster on the night of 2001 July 10.  Both sets utilized
the f/14 tip--tilt system on the Cerro Tololo Interamerican
Observatory (CTIO) 4--m Blanco Telescope using the facility imager
OSIRIS\footnote{OSIRIS is a collaborative project between Ohio State
University and CTIO. Osiris was developed through NSF grants AST
90--16112 and AST 92--18449.}.  Spectroscopic data were acquired with
the Blanco telescope in 2000 May 19, 21 and 22 and 2001 July
11. OSIRIS delivers a plate scale of 0.16$''$ pix$^{-1}$. All basic
data reduction was accomplished using IRAF\footnote{IRAF is
distributed by the National Optical Astronomy Observatories.}. Each
image was flat--fielded using dome flats and then sky subtracted using
a median--combined image of five to six dithered frames. Independent
sky frames were obtained 5--10$'$ south of the G333.1--0.4 cluster for
direct imaging and 1--2$'$ west for spectroscopy.

\subsection{Imaging} 

All images were obtained under photometric conditions. Total exposure
times on 1999's run were 270s, 135s and 81s at
$J$, $H$ and $K$, respectively. The individual $J$, $H$ and $K$
1999's frames were shifted and combined.  These
combined frames have point sources with FWHM of $\approx$ 0.63$"$,
0.54$"$ and 0.56$"$ at $J$, $H$ and $K$, respectively. DoPHOT
\citep{sch93} photometry was performed on the combined images. All
reduction procedures and photometry were performed for each set of
images (1999 and 2001) separately, and the resulting combined
magnitudes were included in a single list. The 1999 images are deeper
than the 2001 images, so we performed photometric completeness tests
and corrections to each set of images independently (see below).

\begin{figure}[finding]
\begin{center}
\includegraphics[totalheight=10.0cm,angle=0]{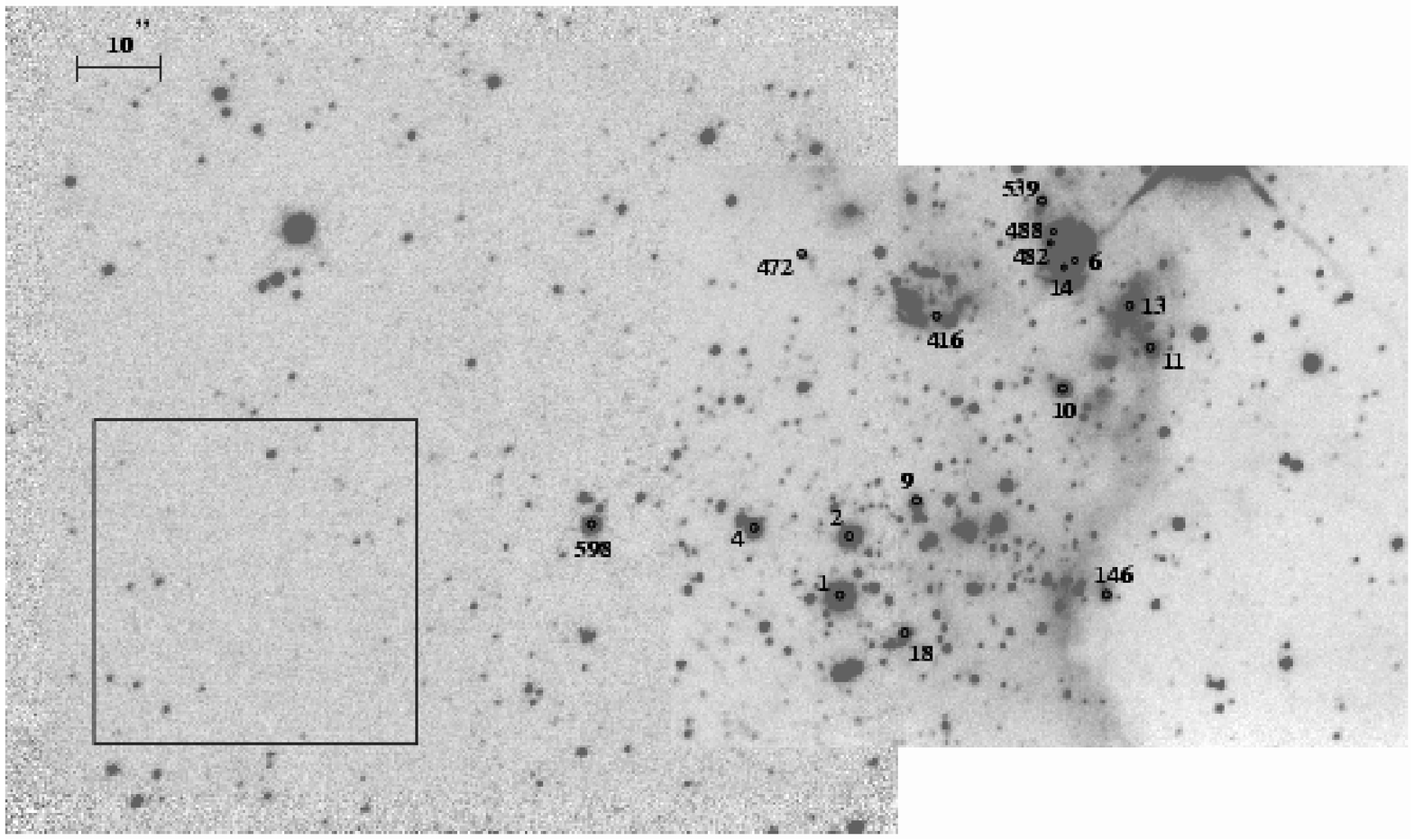}
\end{center}
\caption{Finding chart using a $K$--band image of G333.1--0.4 plus a
background field image 65$''$ east. The square (on the left) indicates
the region that we used to define the background stars (see
text). Object labels refer to the star names for all figures in this
work. North is up and East to the left. The image measures $1'.69$ $\times$
$2'.87$, corresponding to an area of $\approx 4.75 arcmin^{2}$ after
ignoring the two blank strips at top and bottom right. \label{finding}}
\end{figure}

The flux calibration was accomplished with standard star GSPC~S875--C
(also known as [PMK98] 9170) from \citet{per98} which is on the Las
Campanas Observatory photometric system (LCO).  The LCO standards are
essentially on the CIT/CTIO photometric system \citep{eli82}, though
color corrections exist between the two systems for the reddest
stars. No transformation has been derived for OSIRIS and either
CIT/CTIO or LCO systems. Figure~\ref{finding} shows a finding chart
using the $K$--band image made from combination of 1999 and 2001
observations.  The area used to measure the sky counts is shown at the
lower left.

The standard star observations were taken just after the G333.1--0.4
data acquisition and within 0.17 airmass from the target.
The color correction and this remaining airmass
will add uncertainties of the order of 2\% in the worst case
(J--band). No corrections were applied for this small difference in
airmass between target and standard.

Aperture corrections for 16 pixel radius circles were used to put the
instrumental magnitudes on a flux scale. Ten uncrowded stars on the
G333.1--0.4 images were used for this purpose. In order to determine
the zero point for the 2001 images we used stars in common with the
1999 images.

Uncertainties for the $J$, $H$ and $K$ magnitudes in 1999 images
include the formal DoPHOT error added in quadrature to the error in
the mean of the photometric standard and to the uncertainty of the
aperture correction used in transforming from the DoPHOT photometry to
OSIRIS magnitudes. The sums in quadrature of the aperture correction
and standard star uncertainties are $\pm$0.010, $\pm$0.017 and
$\pm$0.044 mag in $J$, $H$ and $K$, respectively. The scatter in the
instrumental magnitudes in the set of stars from the 1999 images used
to calibrate the May 2001 images are $\pm$0.04, $\pm$0.07 and
$\pm$0.06 mag in $J$, $H$ and $K$, respectively. Thus, the errors in
the bright star magnitudes in 2001 images are dominated by this
scatter.  The mean of the final magnitudes errors including all
objects detected are $\pm$0.047, $\pm$0.049 and $\pm$0.078 mag in $J$,
$H$ and $K$ respectively. We adopted an arbitrary cutoff of 0.2 mag
(stars with larger errors were excluded from further analysis).

The completeness of DoPHOT detections was determined through
artificial star experiments. This was accomplished by inserting fake
stars in random positions of the original frame, and then checking how
many times DoPHOT retrieved them. The PSF of the fake star was
determined from an average of real stars found in isolation. Adding a
large number of stars to the real images could effect the
crowding. Instead, we chose to add a small number and repeat the test
many times.  We inserted a total of 24000 stars in the magnitude
interval 8 $\leq$ $K$ $\leq$ 20, corresponding to twenty seven times
the number of real stars recovered in the original DoPHOT run. For
every $\Delta K = 0.5$ we inserted simultaneously 10 stars in the
$K$--band image (240 stars total), and then re--ran DoPHOT to see how
many were recovered. This was repeated 100 times. The completeness of
the sample is defined as the percentage of stars recovered in these
tests. In Figure~\ref{completeness} we present the result of these
experiments -- the photometric completeness. The performance of the
photometry is better than 92\% for a 16th magnitude in the $K$
band. The procedure above was repeated for the $J$ and $H$ bands and
in the both cases the performance of the photometry is better than
92\% for $J < 16.5$ and $H < 16.75$.  The right panel in the
Figure~\ref{completeness} shows the differences between the input
magnitudes of the artificial stars and the output magnitudes of the
artificial stars detected by DoPHOT. Using the magnitude limit (16.0),
the difference between the input and output magnitude of the false
stars is 0.044. This difference is similar to the instrumental
uncertainties given by the DoPHOT. Although we show in
Figure~\ref{completeness} only the results for the 1999 data set
completeness test, the same experiment was performed for the offset
images which have different depth. All data sets were corrected for
completeness separately before constructing the combined luminosity
function.

\subsection{Spectroscopy} 

The $K$--band spectra of three of the brightest stars in the
G333.1--0.4 cluster were obtained: \#1, \#2 and \#4 with OSIRIS. One
dimensional spectra were obtained by extracting and summing the flux
in $\pm$ 2 pixel apertures. The extractions include local background
subtraction from apertures, $\approx$ 1$''$ on either side of the
object. Moreover, we used background apertures in order to subtract the
uniform nebular component of emission from the target spectra. 

Wavelength calibration was accomplished by measuring the position of
bright OH lines from the $K$--band sky spectrum \citep{oli92}. The
spectra were divided by the average continuum of several B9V--type
stars to remove telluric absorption.  The airmass differences between
objects and B9V--type star are $<$ 0.05 and no corrections were
applied for these small differences.

The Br$\gamma$ photospheric feature was removed from the average
B9V--type star spectrum by eye by drawing a line between two continuum
points. Since Br$\gamma$ is free from strong telluric features, it is
sufficient to cut off this line by eye by drawing a line between two
continuum points, to obtain the template for telluric
lines. Br$\gamma$ does play a key role in classification of the
cluster stars. The $K-$band classification scheme for OB stars is
based on faint lines of CIV, HeI, NIII and HeII. Actually, the spectra
of stars in HII region young are often contaminated by the 2.058
$\mu$m HeI and Br$\gamma$ nebular lines, but this is not important,
since these lines are not necessary for classifying O-type stars
\citep{han96}.

The spectral resolution at 2.2 $\mu$m is R $\approx$ 3000 for OSIRIS
and the linear dispersion is $\lambda/pix$ $\approx$ 3.6 $\AA /
pixel$.

\section{Results: Imaging}

The OSIRIS $J$, $H$ and $K$--band images reveal a rich, embedded star
cluster readily seen on the right side of Figure 1, where the stellar
density is higher than the area to the left. We detected a total of
866 stars in the $K$--band image of the cluster and the field located
65$''$ east.  Of those 866 stars, 757 were detected also in the
$H$--band and 343 in all three filters. We have not
detected objects in J or H bands that was not picked up in K with
magnitude errors lesser than the cutoff limit. The image measures
$1'.69$ $\times$ $2'.87$, amounting to an area of $\approx 4.75$
$arcmin^{2}$ -- ignoring the two blank strips at top and bottom right.
A false color image is presented in Figure~\ref{color}, made by
combining the three near infrared images and adopting the colors blue,
green and red, for $J$, $H$, and $K$, respectively. The bluest stars
are likely foreground objects, and the reddest stars are probably
$K$--band excess objects, indicating the presence of hot dust for
objects recently formed in the cluster (background objects seen
through a high column of interstellar dust would also appear red).
The bright ridge of emission that can be seen in the figure crossing
the central region of the cluster in the N--S direction is most likely
due to Br$\gamma$ emission arising from the ionized face of the
molecular cloud from which the cluster has been born. Darker regions
are seen to the west of the emission ridge in
Figure~\ref{color}. This geometry suggests a young
cluster containing massive stars, now in the process of destroying the
local molecular cloud.

The $H - K$ versus $K$ color magnitude diagram (CMD) is
displayed in Figure~\ref{cmd}.  In \S5 we will determine a
spectroscopic parallax of 2.6 kpc for G333.4--0.1. The labels in all
plots refer to the same star as in Figure~\ref{finding}. We can see
two concentrations of objects in the CMD. The first one appears around
$H - K \approx 0.3$, which corresponds to an extinction of $A_{K} =
0.42~mag (A_{V} \approx 4.2~mag)$ using the interstellar reddening
curve of \citep{mat90} (see below).  This sequence represents
foreground stars; the expected extinction for this position along
the Galactic plane is $A_{V} \approx 1.8~mag/kpc$ or $A_{K} \approx
0.18~mag/kpc$ \citep{jk94}.  The second concentration of objects
appears around $H - K = 0.8$ or $A_{K} = 1.22~mag$, probably
indicating the average color of cluster members. A number of stars
display much redder colors, especially the brightest ones in the K-band. 
These objects are located $H - K > 2.0$ or $A_{K}= 3.2$. The dashed vertical
line indicates the position of the theoretical zero age main sequence
(ZAMS; see Table~1 of \citet{blum00}) shifted to 2.6
kpc distance and with interstellar reddening $A_{K} = 0.42~mag$. An
additional local reddening of $A_{K} = 0.80~mag$ results in the ZAMS
position indicated by the vertical solid line.

The $J - H$ versus $H - K$ color--color plot is displayed in
Figure~\ref{ccd}. In that diagram the solid lines, from top to bottom,
indicate interstellar reddening for main sequence M--type
\citep{fro78}, O--type \citep{koo83} and T~Tauri \citep{mey97} stars
(dashed line). The solid vertical line between the main sequence
M--type and O--type lines indicates the position of ZAMS.  Asterisks
indicate $A{_K}$ = 0, 1, 2 and 3 reddening values. Dots are objects
detected in all three filters. The error bars in both figures refers
to the final errors in the magnitudes and colors.

\subsection{Cluster members}

Until now, we have assumed that our sample of stars is composed only
by clusters members (not contaminated by foreground or background
stars). In fact, it is not an easy task to identify these two
populations except through statistical procedures.  All details of our
procedure used to separate the cluster members from projected stars in
the cluster direction can be seen in Figure~\ref{dens}. The left panel
shows the CMD (Figure~\ref{cmd}) binned in intervals of $\Delta K =
1.0$ and $\Delta (H - K) = 1.0$ and containing all stars (clusters
members plus projected stars).  We used the stars in the region
indicated by the square on bottom left in (Figure~\ref{finding}) to
define a field population. In this case, we supposed that there are
only foreground or background stars and no clusters members in this
small area. The star counts inside this square were normalized by the
relative areas projected on the sky and then binned in the same
interval cited above (center panel in Figure~\ref{dens}). The stellar
density in the field (center panel) was then subtracted from the CMD
with all stars (left panel) in bins of magnitude and color intervals,
resulting in a CMD without contamination by projected stars
(statistically; see the right panel). In the case of negative counts
that occur when the field values are bigger than the cluster due to
statistical fluctuations, the counts are added to an adjacent bin that
has a larger number of objects.

This procedure works well for foreground stars, since there are
relatively few stars in the direction of the cluster. For the
background stars, the situation is more complex. However, we believe
the excess of field stars which might contaminate the cluster sample
is relatively small, due to the high obscuration toward the cluster
itself and the large density of cluster stars expected.  Unfortunately
we can not use this procedure to cut off projected stars from our CMD
and/or CCD. However, we can use this result to correct the luminosity
function from contamination by non-cluster members, taking into
account not only the magnitude of the stars but also their colors.
 
\subsection{Reddening and excess emission } 

We estimate the reddening toward the cluster from the extinction law:
$A_K$ $\sim$ 1.6$\times E_{H-K}$ (\citet{ccm89} and \citet{mat90}) and
using an average intrinsic color $H - K$ = $-0.04$ from \citet{koo83}
for OB stars. The \citet{ccm89} extinction law assume $R_{V} =
3.1$. The \citet{ccm89} extinction law is not truly independent of 
environment for $\lambda$ $>$ 0.9 $\mu$m, as pointed out by 
Whitney et al. (2004). However, in the case of the $J$, $H$ and $K$--bands 
(Whitney et al. 2004, Figure~3), differences in the extinction law are small 
enough to be neglected in the present case. The stars
brighter than $K = 14$ have an average observed color of $H - K =
0.8$, corresponding to $A_{K}$ = 1.22 mag ($A_{V}$ $\approx$ 12.2
mag). The interstellar component of the reddening can be separated
from that local to the cluster stars by using the foreground sequence
of stars seen in the CMD diagram (Figure~\ref{ccd}) at $H-K \approx
0.3$, as mentioned in \S3. For G333.1--0.4, the foreground component
is then A$_{K} \approx 0.42$, leaving a local component of A$_{K}
\approx 0.8~$mag.  Regarding the foreground component, as mentioned in
\S3, the value that we have found (A$_{K} \approx 0.42$) to agrees
very well with the value estimated from Jonch-Sorenson, for the
distance of 2.6 kpc (A$_{K} \approx 0.47$).

In order to place the ZAMS in the CMD, we used the $H - K$ colors from
\citet{koo83} and absolute $K$ magnitudes from \citet{blum00}.  The
ZAMS is represented by a vertical solid line in Figure~\ref{cmd},
shifted to $D = 2.6$ kpc and reddened by A$_{K}$ = 0.42 due to the
interstellar component. When adding the average local reddening
(A$_{K} = 0.8)$, the ZAMS line is displaced to the right and down, as
indicated by the dashed lines.  We cannot fix the position of the
ZAMS, since there is a scatter in the reddening. The small group of
relatively bright stars ($K \sim 12$) in between these two lines,
suggests that some of them, the bluer ones, could mark the position of
the ZAMS.

Objects found to the right of the O--type stars reddening line in the
CMD of Figure~\ref{ccd}, shown as a solid diagonal line, have colors
deviating from pure interstellar reddening. This is frequently seen in
young star clusters and is explained by hot dust in the immediate
circumstellar environment. We can estimate a lower limit to the excess
emission in the $K$--band by supposing that the excess at $J$ and $H$
are negligible, and that the intrinsic colors of the embedded stars
are that of OB stars. Indeed, assuming that our sample of stars is
composed of young objects (not contaminated by foreground or
background stars), any OB star would have an intrinsic color in the
range $(H - K)_\circ = 0.0 \pm 0.06$ mag \citep{koo83}.  Let us adopt
for all objects in our sample the intrinsic colors of a B2~V star: $(J
- H)_\circ = -0.09$ and $(H - K)_\circ = -0.04$ \citep{koo83}. The
error in the color index would be smaller than the uncertainty in the
extinction law we are using for the interstellar extinction. From the
difference between the observed $J - H$ and the adopted B2~V intrinsic
$J-H$ color, we obtain the $J$ band extinction by using the extinction
law. In other words, in order to estimate a lower limit to the excess
emission we have assumed the intrinsic colors of all stars that was
detected in $J$, $H$ and $K$--bands to be that of a B2~V
star. Assuming the color $J - H$ is not strongly affected by
circumstellar excess emission, we have determined the line of sight
extinction to each star using the extinction law. We derive the
intrinsic apparent magnitudes based on $A_{J}$ and the intrinsic B2~V
colors. The $K$--band excess emission is then derived as $K_{exc}$ =
$K_\circ$ - ($K$ $-$ A$_K$) or simply, as the difference between the
observed A$_K$ and the A$_K$ estimated from the $J - H$ excess alone.

Our results are displayed in Figure~\ref{Kexc}.  In this plot we have
only included objects with measured $J$, $H$ and $K$ magnitudes. The
solid line indicate $K_{exc}=0$.  Connected solid diamonds refer to
the average value of the $K$ excess in 1 magnitude bins. Dashed lines
indicate 1, 2 and 3 $\sigma$ from the average. Bright objects with
very large excess in the upper right corner of Figure~\ref{Kexc},
cannot be explained by errors in the dereddening procedure because
they have measured $J$, $H$, and $K$. They could represent the
emission from accretion disks around the less massive objects in the
cluster. Objects such as \#488, \#472 and \#416 are well above of the
3 $\sigma$ scatter for otherwise normal stars.

Most of the stars in Figure~\ref{Kexc} have a modest negative excess,
about $-0.2$ magnitude. This negative excess is a consequence of our
assumption that all stars have the intrinsic color of a B2~V type star
and is thus not physical. With reference to Figure~\ref{ccd}, one can
see that any star which lies above the reddening line for a B2~V must,
by definition, have a negative excess under the assumption that the
excess emission is in the $K-$band and the intrinsic photospheric
colors are that for a B2~V star. Our goal is to identify stars with a
significant excess which would cause them to lie to the right of
reddening line in Figure~\ref{ccd}, so this modest negative excess for
``normal'' stars, or stars with a small excess is not important for
our purposes.

In Figure~\ref{Kexc}, the magnitude of the large excess (almost two
mag for object \#6) is in agreement with the values found by
\citet{hil00} for young stars in the Orion cluster. In the following
sections we have only corrected the $K$--excess for stars with
positive excess as determined here, for all others we impose zero
excess emission. Low mass YSOs have typical excess of several 10th's
of a magnitude \citep{hil00}, depending on the age of the cluster.

\subsection{The KLF and the IMF} 

After correcting for non--cluster members, interstellar reddening,
excess emission (a lower limit) and photometric completeness, the
resulting $K$--band luminosity function ($KLF$) is presented in
Figure~\ref{klf}. A linear fit (solid line), excluding deviant
measures by more than $3 \sigma$, has a slope $\alpha = 0.24 \pm
0.02$. A considerably steeper KLF slope was obtained for W42 ($\alpha
= 0.40$) by \citet{blum00} and for NGC3576 ($\alpha = 0.41$) by
\citet{fig02}. A linear fit only using stars measured in all filters
$J$, $H$ and $K$ (dashed line in Figure~\ref{klf}) results in a slope
very close to that found including all stars ($\alpha = 0.26 \pm
0.04$).  The coincidence is not surprising, since the fitting in both
cases is dominated by the objects detected in the three filters.

We can evaluate the stellar masses by using \citet{sch92} models,
assuming that the stars are on the ZAMS instead of the pre--main
sequence.  This is a reasonable approximation for massive members of
such a young cluster.  Stars more massive than about $M = 5 M_\odot$
should be on the ZAMS according to the pre--main sequence (PMS)
evolutionary tracks presented by \citet{sie00}. The main errors in the
stellar masses, given this restriction, will be due to the effects of
circumstellar emission and stellar multiplicity. Our correction to the
excess emission is only a lower limit, since we assumed the excess was
primarily in the $K$ band (but according to Figure~\ref{Kexc}, there
are not many stars with large excess for the higher
masses). \citet{hsvk92} have computed disk reprocessing models which
show the excess in $J$ and $H$ can also be large for disks which
reprocess the central star radiation. An underestimate of the excess
emission will result in an overestimate of the mass for any given star
and the cluster as a whole. The slope of the mass function should be
less effected.  It is difficult to quantify the effect of binarity on
the IMF. If a given source is binary, for example, its combined mass
would be larger than inferred from the luminosity of a ``single'' star
and its combined ionizing flux would be smaller. The cluster total
mass would be underestimated, the number of massive stars and the
ionizing flux would be overestimated. The derived IMF slope would be
flatter than the actual one.

With these limitations in mind, we have transformed the KLF into an
IMF.  Since other authors also do not typically correct for
multiplicity, our results can be inter--compared, as long as this
parameter doesn't change from cluster to cluster. The IMF slope
derived for G333.1--0.4 is $\Gamma = $-$1.1 \pm 0.2$, which is flatter
than Salpeter's slope ($1.5$ $\sigma$ from his value -
\citet{sal55}).  Figure~\ref{imf} shows the binned magnitudes from
Figure~\ref{klf} transformed into masses (triangles) and the fit to
these points considering all objects more massive than $M = 5.0$
$M/M_\odot$ (solid line). The dashed line in the figure indicates a
fit for the case of only those objects more massive than $M = 5.0$
$M/M_\odot$ which have $JHK$ magnitudes measured. The fit ($\Gamma =$
$-$1.0 $\pm$ 0.2) is very close to that derived for all stars. Our
result depends on the calculation of the excess emission which is
uncertain. In the following section, we show that different
assumptions on the excess emission lead to an IMF slope which is
consistent with Salpeter's slope.

\citet{mas95} made a comparison between IMF's of Galactic and LMC OB
associations/clusters and no significant deviation was found from
Salpeter's value. A steeper IMF slope was obtained for the Trapezium
cluster ($\Gamma = -1.43 \pm 0.10$) by \citet{hil00} and for NGC3576
($\Gamma = -1.62 \pm 0.12$) by \citet{fig02}.  Flatter slopes have
been reported only for a few clusters, most notably the Arches and
Quintuplet clusters \citep{fi99a}, both near the Galactic Center.
Flatter slopes may indicate that in the inner Galaxy star forming
regions the relative number of high mass to the low mass stars is
higher than elsewhere in Galaxy. It is also possible that dynamical
effects may be more important in the inner Galaxy. \citet{pz01} have
modeled the Arches cluster data with a normal IMF, but include the
effects of dynamical evolution in the presence of the Galactic center
gravitational potential. They find the observed counts are consistent
with an initial Salpeter--like IMF.

We can determine an approximate lower limit to the total mass of the
cluster by integrating the IMF between $5 < M/M_\odot < 90$.  The
upper integration limit corresponds to the mass of a O3--type star as
given by \citet{blum00}. The integrated cluster mass is $M_{cluster} =
(1.3 \pm 0.5)$ $\times$ $10^{3}$ $M_\odot$.  To the extent that excess
emission is underestimated for these stars, this lower limit to the
cluster mass is overestimated.

The number of Lyman continuum photons derived from the IMF
(Figure~\ref{imf}) was calculated from the contribution of all massive
stars in the cluster. The Lyman continuum flux comes from the
brightest stars which are well above our completeness limit. The
intervals of masses in the IMF have been transformed into Lyman
continuum flux by using results from \citet{vac96} for stars more
massive than 18 $M_\odot$.  We obtain the value for the total NLyc as
$1.9$ $\times$ $10^{50}$ $s^{-1}$. Our new spectrophotometric distance
(see below) and that recently derived from radio observations (see
\S1) put G333.4--0.1 on the near side of the Galactic center;
\citet{smi78} had estimated the NLyC based upon the far side distance.

\subsection{Embedded Young Stellar Objects}

As we can see in Figures~\ref{cmd}, \ref{ccd} and \ref{Kexc}, some
objects in G333.1--0.4 are very bright, display much redder colors and
have excess emission. This is the case for objects such as \#4, \#6,
\#9, \#13, \#14, \#598. In NGC3576 \citep{fig02} we found 4 objects
with photometric evidence for circumstellar emission, and whose
spectra displayed either CO absorption or emission.  Star \#4 in the
present sample also shows such a signature (see the following
section), namely CO band head emission.

We can estimate the intrinsic properties of these YSOs by correcting
for the excess emission and reddening evident in the photometric data
of the color--magnitude diagram. In Table~1, we summarize the
properties of these YSOs with excess emission and extinction as
determined in \S3.2 (only $K$--band excess). In each case, the local
reddening is larger than the mean reddening that we found above for
the cluster. The last two columns in the Table~1 display the spectral
type and corresponding masses of these stars.

Given such evidence for circumstellar disks, we attempt to estimate an
excess emission using \citet{hsvk92} models for reprocessing disks to
get a sense for how much bigger the excess might be compared to that
derived by assuming all the excess is in the $K$--band. By starting
with the maximum excess emission $\Delta K = 4.05$ valid for a
O7--type star (\citet{hsvk92} Table 4: all excess emission values are
for face on line of sight) we derive the corresponding $M_K$.  Using
the ZAMS properties from \citet{blum00} Table 1, we obtain the
corresponding stellar spectral type and mass.  This is generally much
smaller than the O7--type with which we started; the excess is
obviously overestimated and the new luminosity and mass are
underestimated.  From this smaller mass, we use the corresponding
excess emission and find a higher mass and luminosity. We iterate this
procedure until convergence. This was done for all the stars listed in
Table~1, and the results are given in Table~2. The resultant reddening
was determined in the same way as for Table~1, except that now we have
included the excess in the $H-K$ color due to the \citet{hsvk92}
reprocessing disks (nearly constant and equal to 0.5 mag).

The values in Tables~1 \& 2 give a rough indication of the range of
excesses that might be present, though neither is fully correct. The
values in table~2 don't account for accretion luminosity nor for the
inclination of the disk, while those derived only with photometric
data assume negligible excess at $J$ and $H$. A similar procedure
using $JHK$ photometry was adopted for two stars of NGC3576
\citep{fig02}. \citet{B2003} inferred similar spectral types from
mid--infrared imaging techniques.

Although the Table~2 values are not true upper limits, since they
don't account for accretion and may be too extreme if the disk we see
is highly inclined, we can use them as a plausible ``large excess''
case. We used these assumptions on the excess, adopting the masses of
the objects shown in Table~2, in order to compare to those from the
previous IMF estimate (Figure~\ref{imf}). The slope in the case of the
``large excess'' is $\Gamma = -1.3 \pm 0.2$ which is similar than
Salpeter's slope (0.25 $\sigma$ from his value - \citet{sal55}).

New masses, derived from ``large excess'' correction, lead to a NLyc
$=$ $0.2$ $\times$ $10^{50}$s$^{-1}$. This is smaller than the number
of Lyman continuum photons detected from radio observations ($0.6$
$\times$ $10^{50}$s$^{-1}$ at our derived distance). However, the
observed NLyc photons by radio techniques is only a lower limit of
that emitted by the stars, since some of them are destroyed by dust
grains or leaked through directions of low optical depth. In this way,
we can define the lower limit to the IMF slope as $\Gamma < -1.3$.
The integrated cluster mass in this case is $M_{cluster} > 1.0$
$\times$ $10^{3}$ $M_\odot$.

\subsection{G333.1--0.4 \#18}

In Figure~\ref{Kexc} we have only included objects with measured $J$,
$H$ and $K$ magnitudes, but object \#18 was not detected in the J
band, and for this reason needs to be discussed separately. This
object is bright and has the reddest color and largest excess in the
cluster ($K$ = 12.46 and $H - K$ = 5.74). Figure~\ref{obj18} shows
this object in the $J$, $H$ and $K$ bands respectively.  The $K$--band
contours are over--plotted on the $J$ and $H$ images for
comparison. Figure~\ref{obj18} demonstrates that source \#18 is
extremely red and suggests that this object is a deeply buried YSO.

Object \#18 probably is a YSO that is consistent
with an O--type star. A similar object was located in W51 (IRS3 from
\citet{gold94}).  This object has a $K$--excess $>$ 4.05 when using
the model with a reprocessing disk \citep{hsvk92}. Certainly, this is
an object which deserves further study at longer wavelengths and could
aid in trying to understand the processes involved in the formation of
massive stars.

\section{Results: Analysis of Spectra} 

The spectra of sources \#1 and \#2 are shown in Figure~\ref{spec} and
source \#4 is presented in Figure~\ref{yso}. Sources \#1 and \#2 have
been divided by a low--order fit to the continuum after correction for
telluric absorption. The signal--to--noise ratio is S/N $>80$ for each
of these objects. These spectra have been background subtracted with
nearby ($\approx 1''.0$) apertures, though non--uniform extended
emission could affect the resulting He~I and Br$\gamma$ seen in the
stars themselves.

\subsection{O--star spectra}

The spectra of sources \#1 and \#2 may be compared with the $K$--band
spectroscopic standards presented by \citet{han96}. The
features of greatest importance for classification are (vacuum
wavelengths) the CIV triplet at 2.0705, 2.0769 and 2.0842 $\mu$m
(emission), the NIII complex at 2.116 $\mu$m (emission), and HeII at
2.1891 $\mu$m (absorption). The 2.0842 $\mu$m line of CIV is typically
weak and seen only in very high signal--to--noise spectra
\citep{han96}. The present classification system laid out by \citet{han96}
does not have strong luminosity--class indicators. Still,
the HeI (2.0581 $\mu$m) and Br$\gamma$ (2.1661 $\mu$m) features can be
used to approximately distinguish between dwarfs plus giants on the
one hand, and supergiants on the other.  Generally strong absorption
in Br$\gamma$ is expected for dwarfs and giant stars and weak
absorption or emission for supergiants.

The presence of NIII and HeII in the spectrum of source \#1 (see
Figure~\ref{spec}) leaves no doubt that this is a O--type star. The
CIV emission places the source \#1 in the kO5--O6 subclass. The
apparent absorption feature at the position of Br$\gamma$ and of HeI
suggest that source \#1 probably is a dwarf or a giant star.  The
closest match in \citet{han96} atlas is star HD 93130
classified as O6III(f).  However, it is very difficult to be sure
about the exact luminosity class. In our earlier work, we adopted a
ZAMS classification due to the presence of massive YSOs in the
cluster. Following the same reasoning, we classify object \#1 as O6V.

The spectrum of source \#2 shows HeI at 2.0581 $\mu$m (emission), HeI
at 2.1137 $\mu$m (weak absorption), NIII at 2.116 $\mu$m (emission)
and Br$\gamma$ (2.1661 $\mu$m) in absorption.  The presence of HeI and
the absence of the CIV triplet indicate that this star is cooler than
source \#1 and a comparison with the Hanson's standards gives an O8V
type star. In any way, as we said in \S2.2, the spectra 
of stars in HII region are often contaminated by the 2.058 $\mu$m HeI and 
Br$\gamma$ nebular lines. In the source \#1 it is not so critical 
but, in the case of source \#2 it will deserves better
$S/N$ spectra to say definitively its spectral type. 

\subsection{G333.1--0.4 \#4}

The spectrum of G333.1--0.4 \#4 is shown in Figure~\ref{yso}. This
object doesn't show photospheric lines, indicating that it is still
(at least partially) enshrouded in its birth cocoon. This is
corroborated by the excess emission derived from photometry (see
Table~1 and Figure~\ref{ccd}).

The CO band head at 2.2935 $\mu$m appears in emission, and it is
usually attributed to warm ($> 1000 K$), very dense ($\rho \approx
10^{10}$ $cm^{-3}$) circumstellar material near the star
\citep{scov83,carr89}. However, a variety of mechanisms and models
have been proposed to explain the origin of CO emission in YSOs. These
include circumstellar disks, stellar or disk winds, magnetic accretion
mechanisms such as funnel flows, and inner disk instabilities similar
to those which have been observed in FU Orionis--like objects and
T~Tauri stars in a phase of disk accretion
\citep{carr89,carr93,chan93,bis97}. \citet{han97} reported the
presence of CO in emission in several massive stars in M17 and
\citet{fig02} also found CO emission in a massive YSO in NGC3576 as
mentioned previously. A high resolution spectrum of source \#4 is
presented by \cite{blum04} who show that the emission is consistent
with a disk origin.
       
\section{Distance Determination} 

In the previous section we classified the spectra of two brightest
stars in G333.1--0.4 as O type stars (O6 and O8). We can now estimate
the distance to G333.1--0.4 by using the spectroscopic and photometric
results. We compute distances assuming the O stars shown in the
Figure~\ref{spec} are zero--age main--sequence (ZAMS) or in the dwarf
luminosity class (i.e., hydrogen burning). The spectral type in each
case is assumed to be O6 (star \#1) and O8 (star \#2).  For the ZAMS
case, the $M_{K}$ is taken from \citet{blum00}. For the dwarf case, the
distance is determined using the $M_V$ given by \citet{vac96} and
$V-K$ from \citet{koo83}. The distance estimates are shown in
Table~3. For the derived spectral types, we obtain distances of $2.6
\pm 0.4$ and $3.5 \pm 0.7$ kpc for the ZAMS and dwarf cases,
respectively. The former value is to be preferred given the presence
of massive YSOs in the cluster. The uncertainty quoted for the mean
distance is the standard deviation in the mean of the individual
distances added in quadrature to the uncertainty in $A_K$ (250 -- 500
pc).

Our distance estimates are in close agreement with the near distance
given by \citet{vil00}: 2.8 kpc.  Their distance was
obtained by the radio recombination line velocity and Galactic
rotation model. \citet{smi78} estimated the $L_{yc}$ luminosity
of G333.1--0.4 to be $10.8$ $\times$ $10^{50}$ s$^{-1}$ assuming a far
kinematic distance (10.7 kpc). Adopting our mean value of $2.6$ kpc, as
indicated by the spectroscopic parallax, considerably reduces the
expected ionizing flux from the radio continuum measurements to $0.6$
$\times$ 10$^{50}$ s$^{-1}$. This value is about three times smaller than the 
value of $1.9$ $\times$ 10$^{50}$ s$^{-1}$ derived from counting the 
individual stars and using the mass function (see \S3.3 above).

\section{Discussion and Summary}

\begin{figure}[color]
\begin{center}
\includegraphics[totalheight=13.0cm,angle=0]{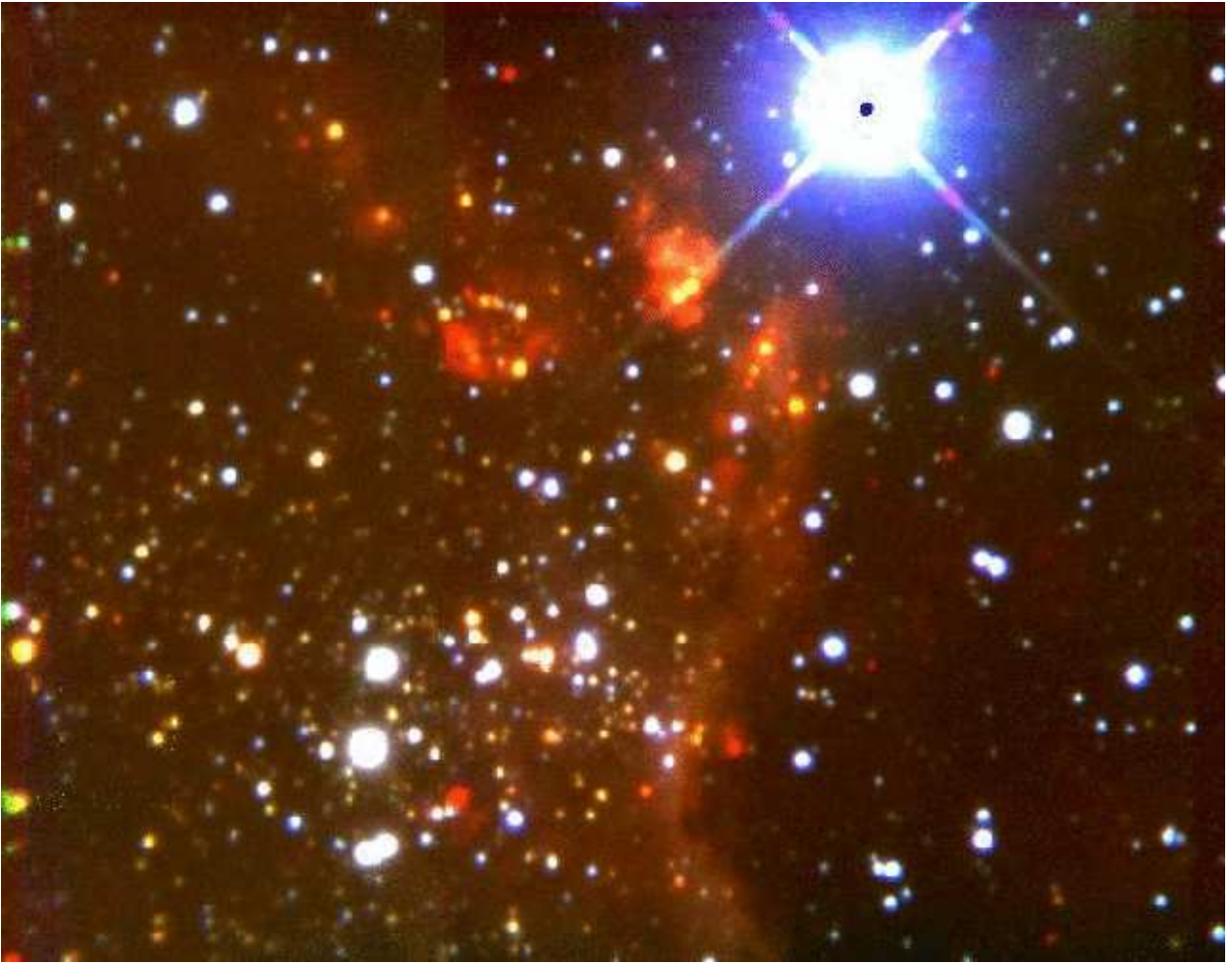}
\end{center}
\caption{False color image of G333.1--0.4: $J$ is blue, $H$ is green
and $K$ is red. The coordinates of the center of the image are RA
(2000) = 16h21m03.3s and Dec. = $-$50$^{0}$36$'$19$''$ and the size of
the image is 1.9$'$ $\times$ 1.7$'$ (plate scale = 0.16"/pixel). North is
up and East to the left.\label{color}}
\end{figure}

\begin{figure}[completeness]
\begin{center}
\includegraphics[totalheight=6.0cm,angle=0]{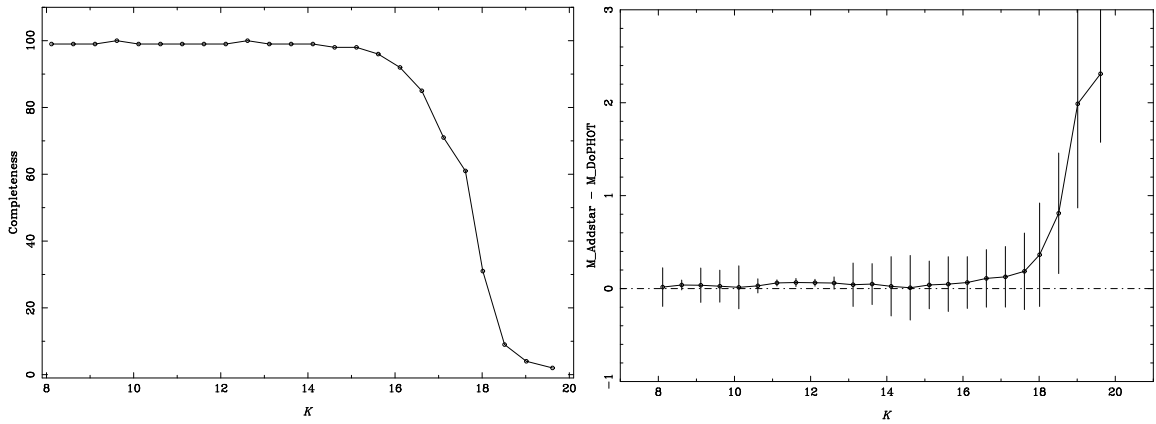}
\end{center}
\caption{Derived completeness for the cluster photometry. The {\it
left} panel shows the completeness (in percent detection) as derived
from artificial star experiments (see text). The {\it right} panel
displays the differences between the input magnitudes of the
artificial stars and the output magnitudes as detected by DoPHOT (see
text). \label{completeness}}
\end{figure}

\begin{figure}[cmd]
\includegraphics[totalheight=17.0cm,angle=-90]{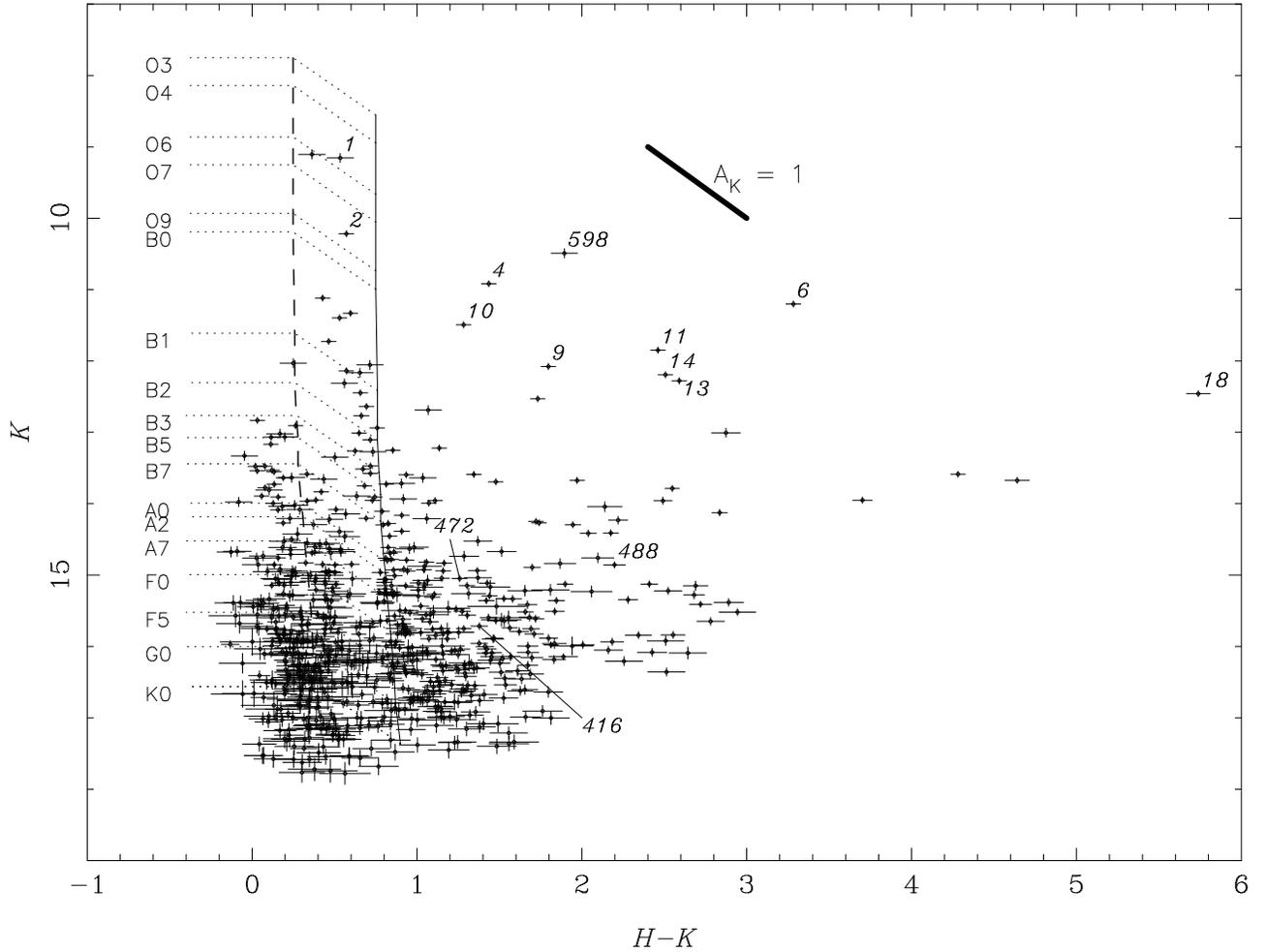}
\caption{$K$ vs $H - K$ color--magnitude diagram (CMD).  The left {\it
dashed} line indicates the position of the theoretical ZAMS shifted to
2.6 kpc and with interstellar reddening $A_{K} = 0.42~mag$.  An
additional ``cluster'' reddening component of $A_{K} = 0.80~mag$
($A_{K total} = 1.22~mag$) results in the ZAMS position indicated by
the vertical {\it solid} line. Object number labels are the same as in
Figure~\ref{finding}.\label{cmd}}
\end{figure}

\begin{figure}[ccd]
\begin{center}
\includegraphics[totalheight=17.0cm,angle=-90]{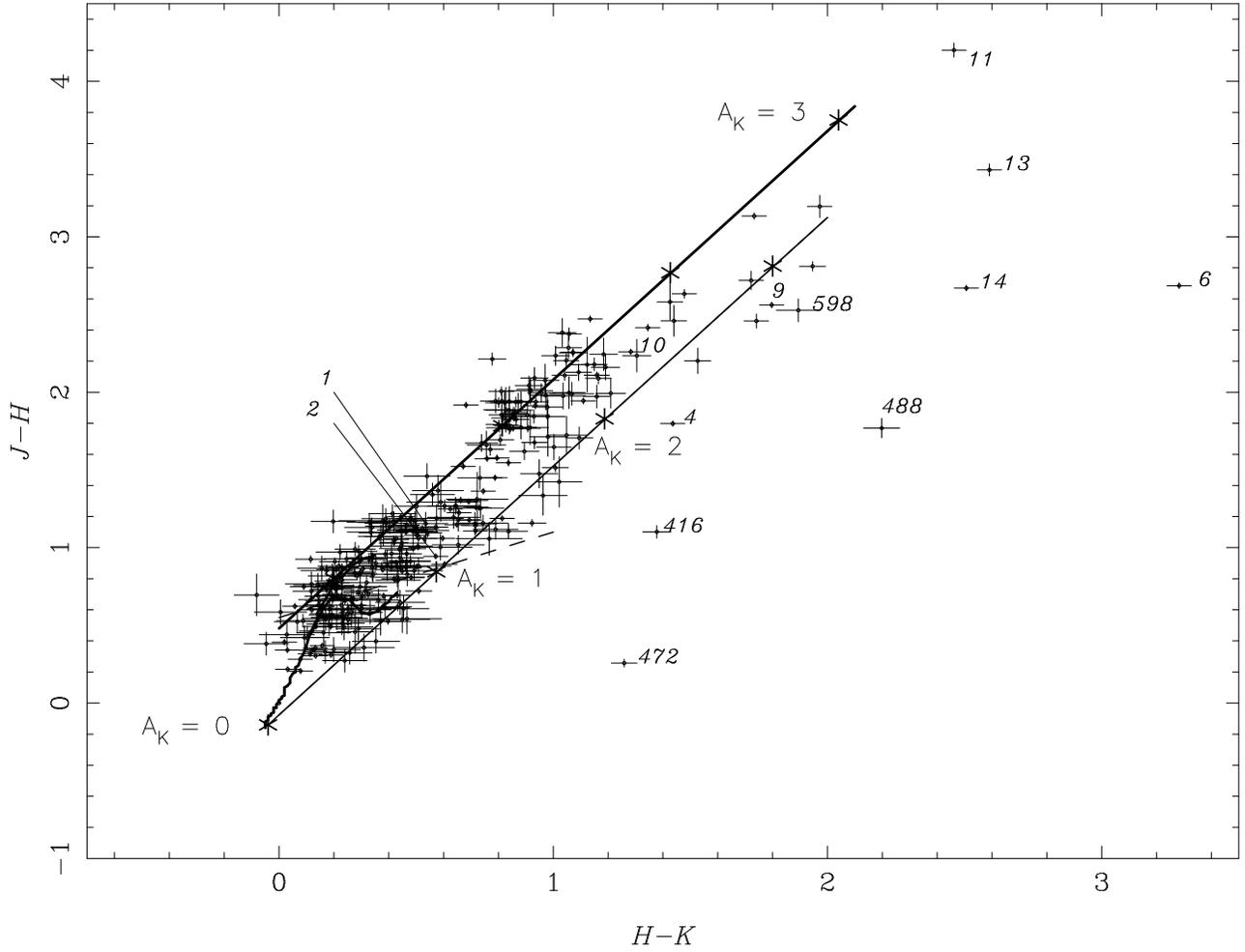}
\end{center}
\caption{$J - H$ vs $H - K$ color--color plots showing the reddening
line of M--type stars ({\it heavy solid} line), O--type stars ({\it
solid} line) and T Tauri stars ({\it dashed} line). {\it Dots} refer
to stars detected in the three filters. The {\it asterisks} indicate
the corresponding $A_{K}$ along the reddening vector.
\label{ccd}}
\end{figure}

\begin{figure}[dens]
\begin{center}
\includegraphics[totalheight=7.5cm,angle=0]{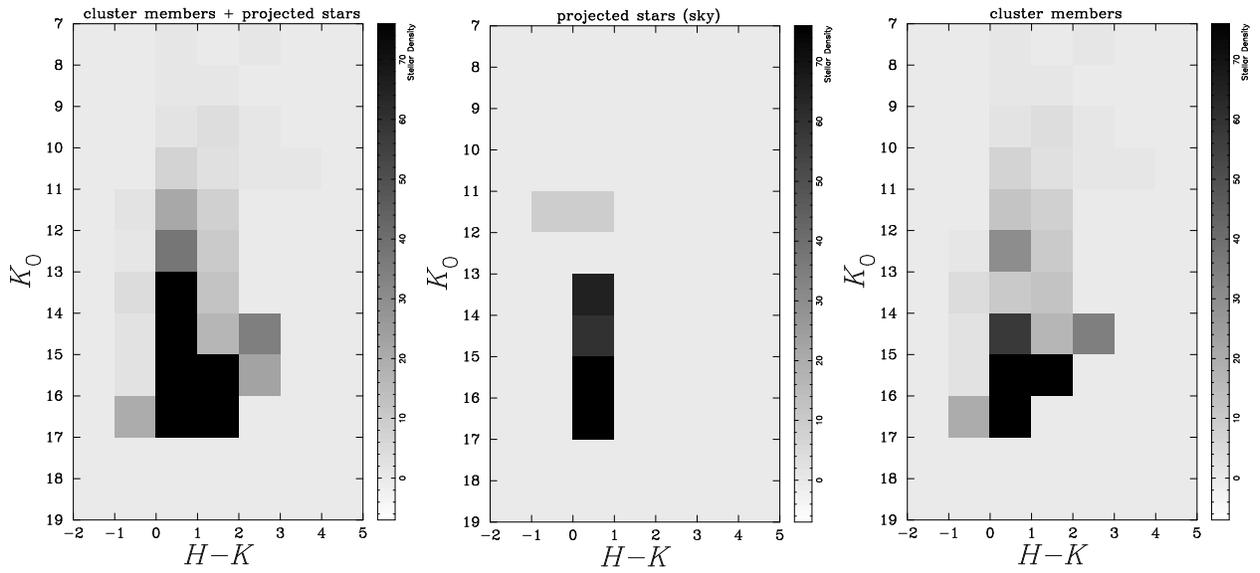}
\end{center}
\caption{Color--Magnitude diagram binned in intervals of $\Delta K =
1.0$ and $\Delta (H - K) = 1.0$ in order to separate the cluster
members from projected stars in the cluster direction. 
The complete CMD taking into account all stars 
is given in the {\it left} panel and the field CMD (represented by 
the square in Figure~\ref{finding}) in the {\it center}
panel. The star counts were normalized by the relative areas
projected on the sky. The {\it right} panel shows the cluster CMD
obtained as the difference between the complete and field CMDs.
\label{dens}}
\end{figure}

\begin{figure}[Kexc]
\includegraphics[totalheight=17.0cm,angle=-90]{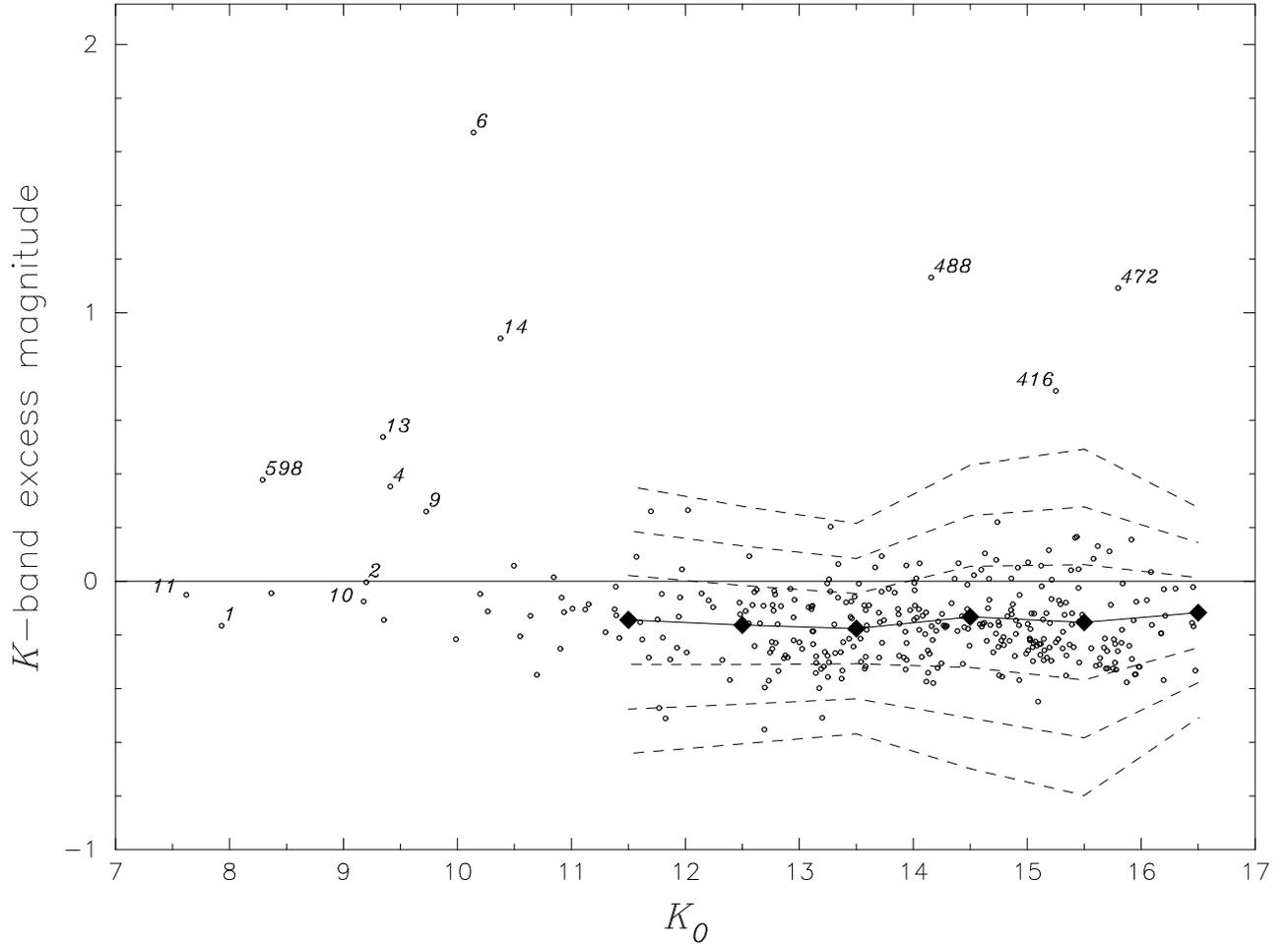}
\caption{Excess emission as a function of dereddened $K$--band
magnitude ($K_{\circ}$). In this plot we have only included objects
with measured $J$, $H$ and $K$ magnitudes (dots). The solid line
indicate $K_{exc}=0$. Connected solid diamonds refer to the average
value of the $K$ excess in one magnitude bins. Dashed lines indicate one, 
two, and three $\sigma$ from the average. Very positive values
represent circumstellar emission. \label{Kexc}}
\end{figure}

\begin{figure}[klf]
\includegraphics[totalheight=17.0cm,angle=-90]{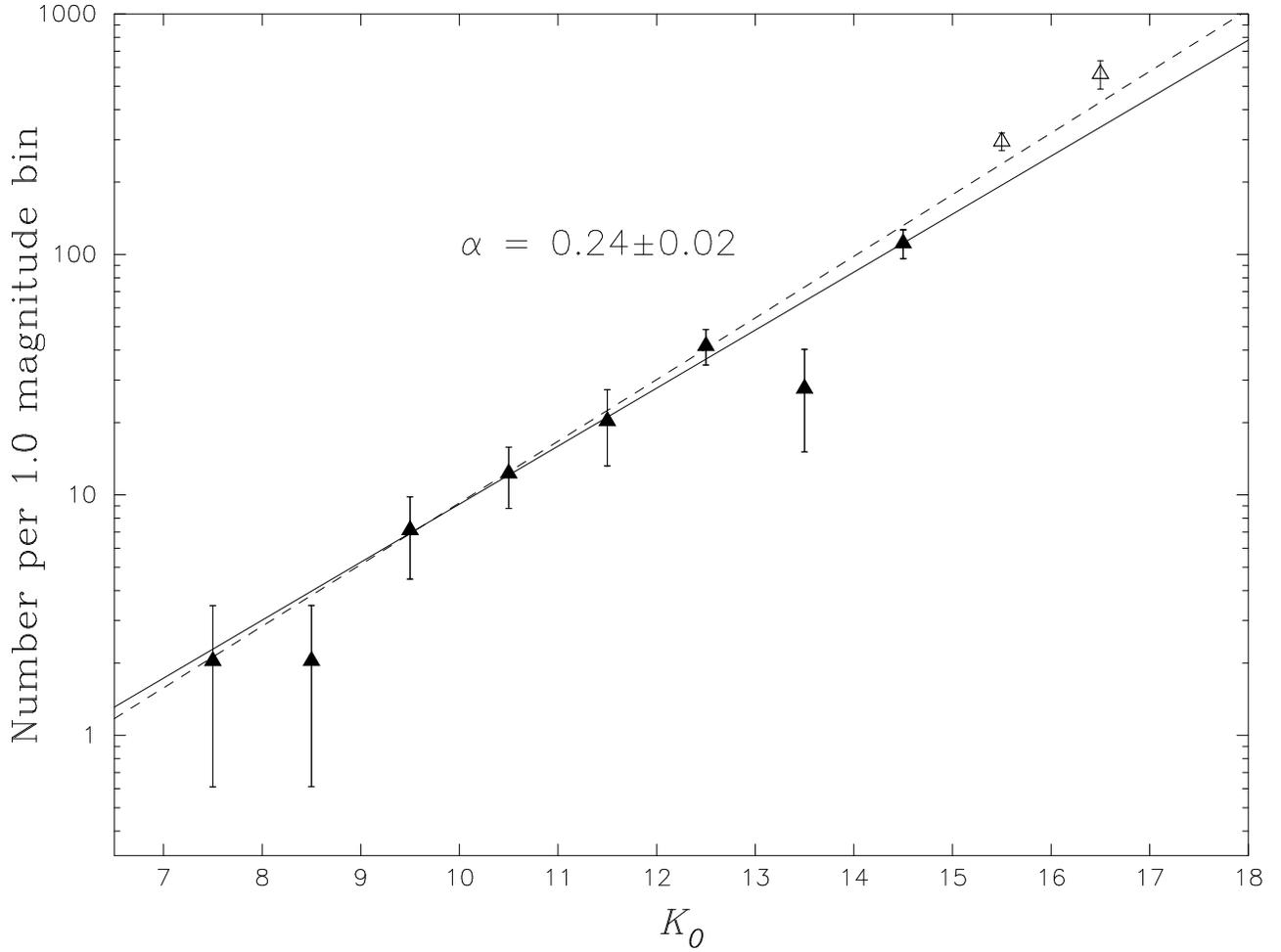}
\caption{The $K$--band luminosity function of the cluster (cluster members 
CMD; see Figure~\ref{dens} right panel), corrected for sample
incompleteness. $K_{\circ}$ is dereddened and corrected for excess
emission; see text.  A linear fit for stars measured in all filters
(dashed line) results in a slope very similar to that found using all
detected objects ({\it solid} line). Open triangles refer to bins 
that were not considered on the fit.\label{klf}}
\end{figure}

\begin{figure}[imf]
\includegraphics[totalheight=17.0cm,angle=-90]{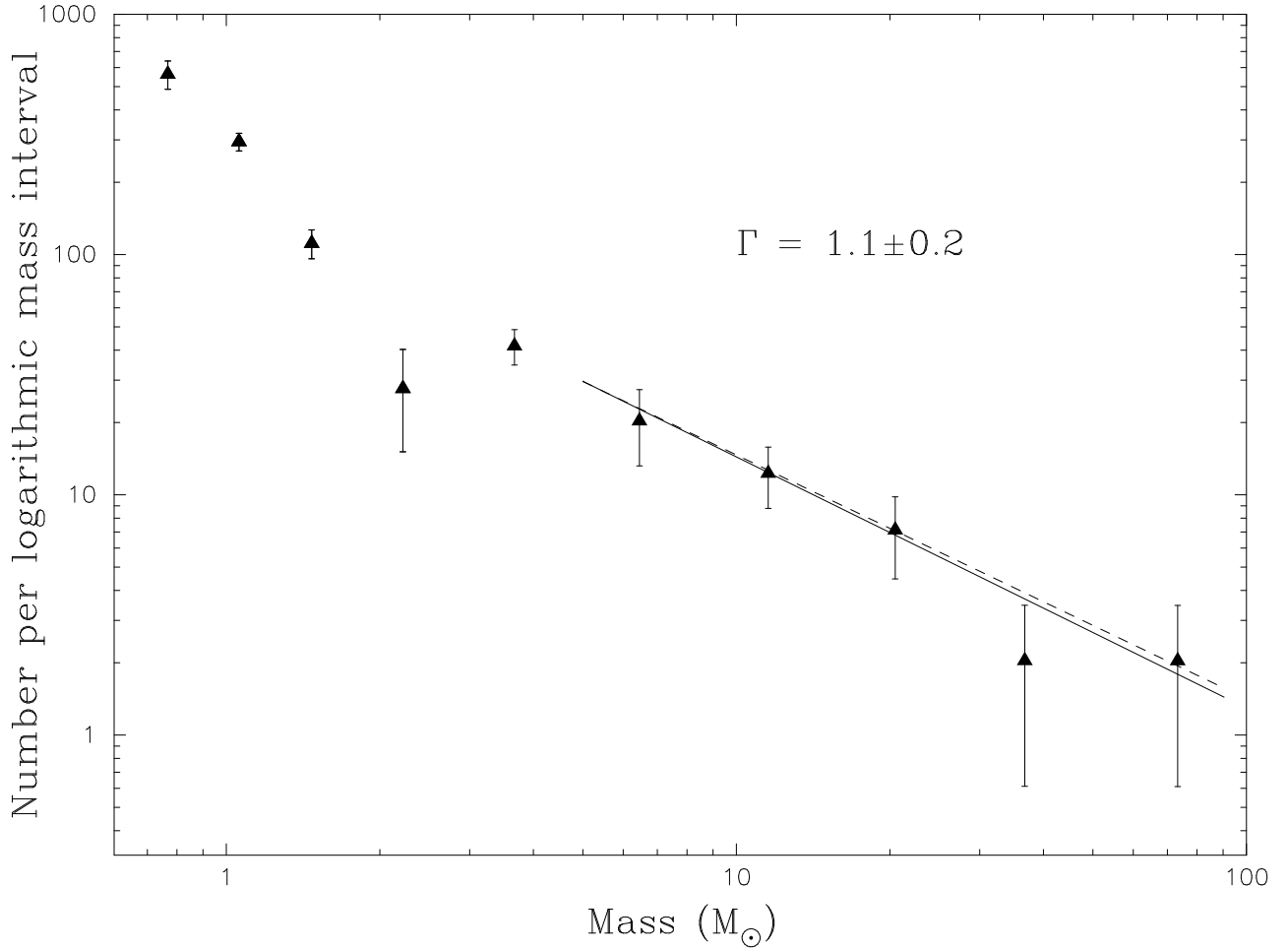}
\caption{The IMF of the cluster members applying \citet{sch92} models
to the corrected $K$--band luminosity function from
Figure~\ref{klf}. Using all stars in the sample, the best fit slope is
$\Gamma =$ $-$1.1 $\pm$ 0.2 ({\it solid} line), consistent with
\citet{sal55}. Fitting only stars measured in all three filters ({\it
dashed} line) results in a similar slope ($\Gamma =$ $-$1.0 $\pm$ 0.2). 
Only stars with $M$ $>$ 5 $M_\odot$ are used in determining the mass function slope; see 
text.\label{imf}}
\end{figure}

\begin{figure}[obj18]
\begin{center}
\includegraphics[totalheight=12.0cm,angle=0]{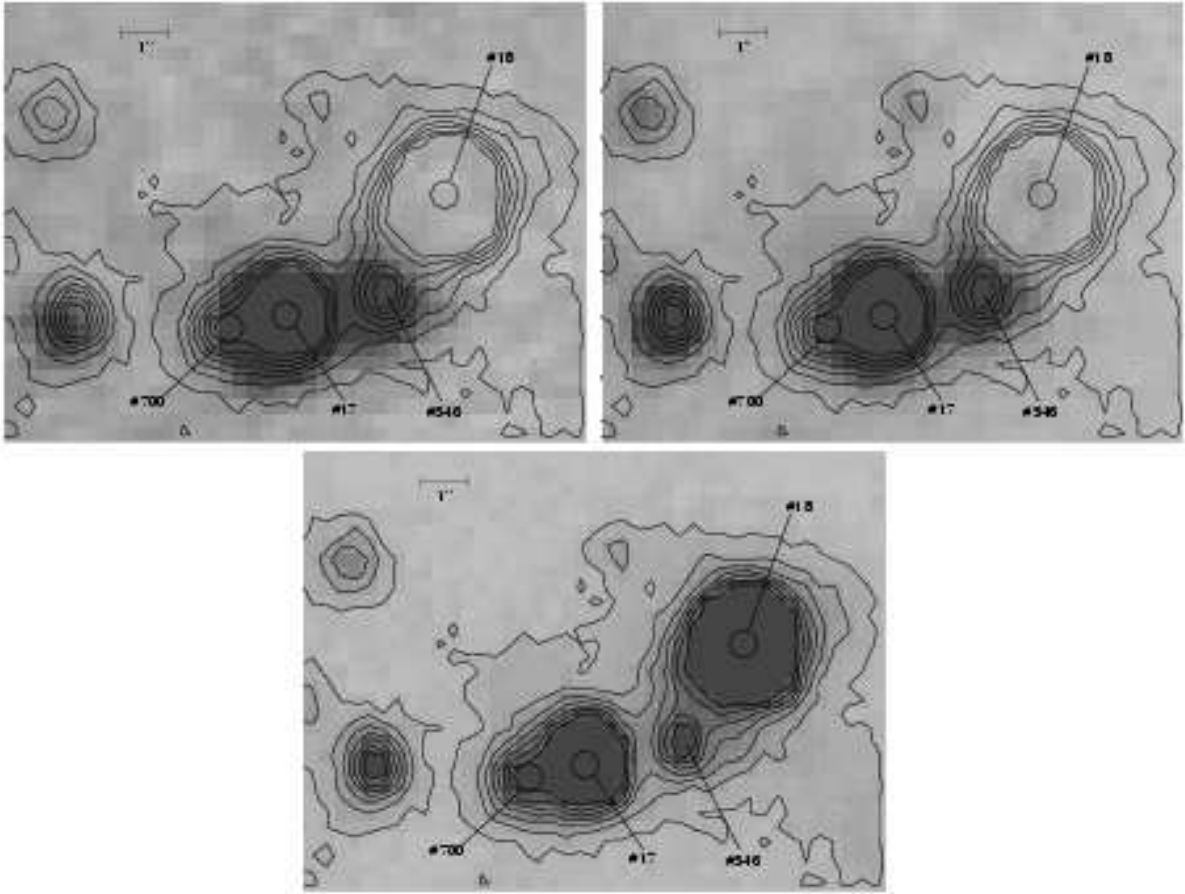}
\end{center}
\caption{$J$, $H$ and $K$ bands images of G333.1--0.4\#18 with
contours overplotted to highlight the difference in flux between
longer and shorter wavelengths. Object \#18 is located at $\alpha$ $=$
16h21m02.62s and $\delta$ $=$ $-$50$^{\deg}$35$'$54.9$''$.
\label{obj18}}
\end{figure}

\begin{figure}[spec]
\includegraphics[totalheight=7.0cm,angle=0]{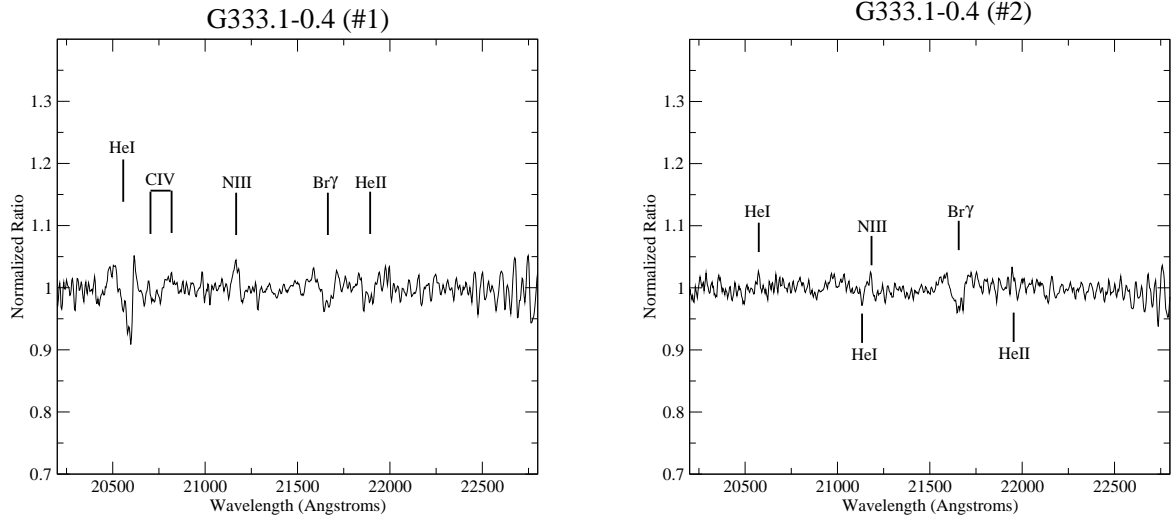}
\caption{$K$--band spectra for the two brightest stars in the G333.1--0.4
cluster: \#1 (O6) and \#2 (O8V). The two pixel resolution gives
$\lambda/\Delta\lambda \approx 3000$. Spectra were summed in apertures
$0''.64$ wide by a slit width of $0''.48$ and include background
subtraction from apertures centered $\le 1''.0$ on either side of the
object. Each spectrum has been normalized by a low--order fit to the
continuum (after correction for telluric absorption).
The spectra are often contaminated by the 2.058 $\mu$m 
HeI and Br$\gamma$ nebular lines.\label{spec}}
\end{figure}

\begin{figure}[spec2]
\begin{center}
\includegraphics[totalheight=8.0cm,angle=0]{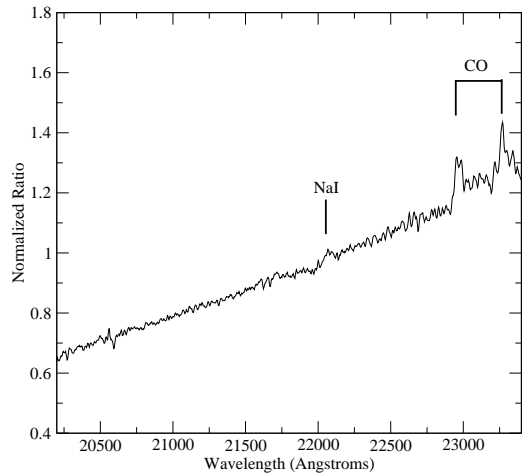}
\end{center}
\caption{$K$--band spectra for G333.1--0.4 \#4 displaying featureless
continua superimposed on CO emission. A NaI line is present at 2.207
$\mu$m.\label{yso}}
\end{figure}

We have presented deep $J$, $H$ and $K$ images of the stellar cluster
in G333.1--0.4 (Figure~\ref{color}) and $K$--band spectra for three
cluster members. Two of them have classic O star absorption lines. The
spectrum of G333.1--0.4 \#4 (Figure~\ref{yso}) doesn't show
photospheric lines but rather CO emission. These features indicate
that it is still enshrouded in its birth cocoon and is perhaps
surrounded by a circumstellar disk. The $K$--band excess emission
displayed by objects \#4, \#6, \#9, \#13, \#14, \#18, \#483, \#488 and
\#158 is similar to objects found in other GHII regions. These objects
appear to be still heavily enshrouded by circumstellar envelopes
and/or disks.  Object \#18 is an extremely buried YSO, and it deserves
followup observations at longer wavelengths to further investigate its
nature.

The KLF and IMF were computed and compared with those of other massive
star clusters. The slope of the $K$--band luminosity function ($\alpha
= 0.24 \pm 0.02$) is similar to that found in other young clusters,
and the IMF slope of the cluster, $-1.3 < \Gamma < -1.1$, is 
consistent with Salpeter's value within 1.25 $\sigma$.

Spectral types of the newly identified O stars and the photometry
presented here constrain the distance to G333.1--0.4, which was
uncertain from earlier radio observations. Our measurements 
break the ambiguity in the distance determinations from radio 
techniques. Our value, 2.6 $\pm$ 0.4 kpc, is consistent with the 
lower distance determined by \citet{vil00} and implies in 
NLyc $=$ 0.6 $\times$ 10$^{50}$ s$^{-1}$, what is considerably 
lower than that adopted by \citet{smi78}. The number of Lyman
continuum photons derived from the contribution of all massive
stars in the cluster is $0.2$ $\times$ $10^{50}$ $s^{-1}$
$< NLyc < 1.9$ $\times$ $10^{50}$ $s^{-1}$. 
The integrated cluster mass is $1.0$ $\times$ $10^{3}$ $M_\odot < M_{cluster} 
< 1.3$ $\times$ $10^{3}$ $M_\odot$.

EF and AD thank FAPESP and CNPq for support. PSC appreciates
continuing support from the National Science Foundation. We thank an
anonymous referee for the careful reading of this paper and for the
useful comments and suggestions which have resulted in a much improved
version. Thanks also go to D. Figer for his helpful comments.

\clearpage


\clearpage

\clearpage 

\clearpage 

\clearpage 

\clearpage 

\clearpage 

\clearpage 

\clearpage 

\clearpage 

\clearpage

\clearpage

\pagestyle{empty}
\begin{deluxetable}{lrcrcccc}
\tablecaption{YSO Properties from Photometric Data \label{YSOs}} 
\tablewidth{0pt}
\tablehead{ \colhead{ID} & \colhead{$J-H$} &
\colhead{$H-K$} & \colhead{$K$} &
$K$--excess\tablenotemark{a} & $A_{K}$\tablenotemark{b} & SP Type\tablenotemark{c} & Mass\tablenotemark{c}}
\startdata
\#4    &  1.80 $\pm$ 0.02 & 1.44 $\pm$ 0.04  &  10.92 $\pm$ 0.04 & 0.35  &  2.37 & O7.5 &   29	  \\
\#6    &  2.69 $\pm$ 0.02 & 3.28 $\pm$ 0.04  &  11.20 $\pm$ 0.04 & 1.67  &  5.31 & O4   &   70	  \\
\#9    &  2.56 $\pm$ 0.02 & 1.80 $\pm$ 0.04  &  12.08 $\pm$ 0.04 & 0.26  &  2.94 & O9   &   22	  \\
\#10   &  2.26 $\pm$ 0.02 & 1.28 $\pm$ 0.05  &  11.49 $\pm$ 0.04 & -0.08 &  2.11 & O8.5 &   23	  \\
\#11   &  4.20 $\pm$ 0.05 & 2.46 $\pm$ 0.05  &  11.85 $\pm$ 0.04 & -0.05 &  4.00 & O4.5 &   58	  \\
\#13   &  3.43 $\pm$ 0.04 & 2.59 $\pm$ 0.04  &  12.28 $\pm$ 0.04 & 0.54  &  4.21 & O6.5 &   34	  \\
\#14   &  2.67 $\pm$ 0.02 & 2.51 $\pm$ 0.04  &  12.19 $\pm$ 0.04 & 0.91  &  4.08 & O8   &   27	  \\
\#416  &  1.10 $\pm$ 0.04 & 1.38 $\pm$ 0.05  &  15.72 $\pm$ 0.05 & 0.71  &  2.27 & A7   &   1.7   \\
\#472  &  0.26 $\pm$ 0.03 & 1.26 $\pm$ 0.05  &  15.05 $\pm$ 0.05 & 1.09  &  2.08 & A5   &   1.7   \\
\#488  &  1.77 $\pm$ 0.06 & 2.20 $\pm$ 0.07  &  14.86 $\pm$ 0.04 & 1.13  &  3.58 & B4   &   4	  \\
\#598  &  2.53 $\pm$ 0.07 & 1.89 $\pm$ 0.08  &  10.49 $\pm$ 0.06 & 0.38  &  3.09 & O4.5 &   59	  \\
\enddata 

\tablenotetext{a}{Assuming \citet{koo83} color for normal stars and
that all excess emission is in the $K-$band.}

\tablenotetext{b}{Derived reddening after correcting for the excess in
the $K$ band; see \S3.4}

\tablenotetext{c}{Properties derived from corrected $K$ magnitudes,
assumed intrinsic colors, and properties of ZAMS or typical OB stars.
We obtain the stellar spectral type and mass using the ZAMS properties
from \citet{blum00}; see text.}

\end{deluxetable}

\pagestyle{empty}
\begin{deluxetable}{lcrcc}
\tablecaption{YSO Properties Using Reprocessing Disk Models  
\label{YSO2s}} \tablewidth{0pt}
\tablehead{ \colhead{ID} & 
$K$--excess\tablenotemark{a} & 
$A_{K}$\tablenotemark{b} & 
SP Type\tablenotemark{c} & 
Mass\tablenotemark{c}}

\startdata
\#4    & 2.9 	&  1.57  & B4	&  4	\\
\#6    & 3.5 	&  4.51  & B0.5 &  14	\\
\#9    & 2.8 	&  2.14  & B6	&  3	\\
\#10   & 2.4 	&  1.31  & B5	&  4	\\	   
\#11   & 3.0 	&  3.2	 & B2	&  6	\\
\#13   & 3.0 	&  3.41  & B2	&  5	\\
\#14   & 3.0 	&  3.28  & B2	&  5	\\
\#416  & 1.1 	&  1.47  & G0	&  1	\\
\#472  & 1.3 	&  1.28  & F5	&  1.2  \\
\#488  & 1.7 	&  2.78  & A2	&  2	\\
\#598  & 3.2 	&  2.29  & B1	&  7	\\
\enddata 

\tablenotetext{a}{Excess emission resulting from a face--on
reprocessing disk with a central source corresponding to the spectral
type listed in column 4 \citep{hsvk92}.}

\tablenotetext{b}{Resultant extinction after correcting for the excess
$H-K$ using \citet{hsvk92} models for reprocessing disks.  The $H-K$
excess is nearly constant and equal to $\approx$ 0.5 mag; see \S3.4.}

\tablenotetext{c}{Spectral type and mass obtained from the observed
photometric data, extinction correction, excess emission models of
\citet{hsvk92}, and ZAMS properties from \citet{blum00}.}

\end{deluxetable}

\pagestyle{empty}
\begin{deluxetable}{lrcccc}
\tablecaption{O--type Stars Properties \label{otype}} \tablewidth{0pt}
\tablehead{ \colhead{ID} & \colhead{$K$\tablenotemark{a}} &
\colhead{$H-K$\tablenotemark{a}} &
$A_{K}$\tablenotemark{b} & $D_{ZAMS}$\tablenotemark{c} & 
$D_{V}$\tablenotemark{c}}
\startdata
\#1	&   9.16 $\pm$ 0.06 & 0.54 $\pm$ 0.08 &  0.94            & 2.4		  & 3.0		      \\
\#2	&  10.22 $\pm$ 0.04 & 0.57 $\pm$ 0.04 &  0.98            & 2.7		  & 3.9	              \\
Average & 	            & 	              &  0.96 $\pm$ 0.15 & 2.6 $\pm$ 0.4  & 3.5 $\pm$ 0.7     \\
\enddata
\tablenotetext{a}{Uncertainty in photometry is the sum in quadrature of the 
photometric uncertainty plus the PSF--fitting uncertainty; see \S2.)}

\tablenotetext{b}{The uncertainty in $A_{K}$ is dominated by the
variation in the power--law exponent of the interstellar
extinction law \citep[$\pm0.16$ - ][]{ccm89}; see § 3.2. The uncertainty in mean AK is the sum 
in quadrature of the standard deviation in the mean plus the (0.15 mag) 
systematic uncertainty due to the extinction law.}

\tablenotetext{c}{Distance estimates assuming mean ZAMS and dwarf (V)
luminosities, see text. The uncertainty in the distance is taken as the sum 
in quadrature of the standard deviation in the mean of the individual estimates 
plus a component (250 -- 500 pc) due to the 
systematic uncertainty in AK.}

\end{deluxetable}
\end{document}